\documentclass[twocolumn,aps,pre,showpacs,amsmath,amsfonts,amssymb,floatfix,pre]{revtex4}
\usepackage{graphicx}
\usepackage{dcolumn}
\usepackage{bm}

\begin{document}

\title{Spectral analysis of deformed random networks}

\author{Sarika Jalan}
\email{physarik@nus.edu.sg}
\thanks{presently at: National University of Singapore, 2 Science Drive 3, Singapore 117542}
\affiliation{Max-Planck Institute for the
Physics of Complex Systems, N\"{o}thnitzerstr. 38, D-01187 Dresden, Germany}

\begin{abstract}
We study spectral behavior of sparsely connected random networks under the random
matrix framework. Sub-networks without any connection among them form a network having 
perfect community structure. As connections among the sub-networks are introduced, 
the spacing distribution shows a transition from the Poisson statistics to the Gaussian orthogonal 
ensemble statistics of random matrix theory. The eigenvalue density distribution shows a transition to 
the Wigner's semicircular behavior for a completely deformed network. 
The range for which spectral rigidity, measured by the Dyson-Mehta $\Delta_3$ statistics,  
follows the Gaussian orthogonal ensemble statistics depends 
upon the deformation of the network from the perfect community structure. The spacing distribution
is particularly useful to track very slight deformations of the network from a perfect community structure, 
whereas the density distribution and the $\Delta_3$ statistics remain identical to the undeformed
network. On the other hand the $\Delta_3$ statistics is useful for the larger deformation strengths.
Finally, we analyze the spectrum of a protein-protein interaction network for 
Helicobacter, and compare the spectral behavior with those of the model networks.
\end{abstract}
\maketitle

\section{Introduction}
The network concept has been gaining recognition as a fundamental tool in
understanding the dynamical behavior and the response of real systems
from different fields such as biology, social systems, technological systems. 
Examples of biological systems include food-web,
nervous system, cellular metabolism, protein-protein interaction network, gene 
regulatory networks; social systems include scientific collaboration, citation, 
linguistic networks, and technological systems include internet, power-grid  
\cite{rev-network}. Many of these networks have been shown to have universal
structural properties, such as degree distribution following a power law, small diameter,
large clustering coefficient, existence of communities \cite{rev-network,SW,BA,Newman}.

Different network models have been proposed and investigated
in detail to understand systems having an underlying network structure
\cite{rev-network,SW,BA,Amaral}. 
These models concentrate to capture one or more structural properties of the networks mentioned above
\cite{SW,BA,rev-network}. Apart from these direct measurements
of structural properties, network spectra are also useful to understand
various properties of the underlying system.  Eigenvalues of the adjacency 
matrix of networks form what are called network spectra, and provide information about
some basic topological properties of the underlying network \cite{spectrum,handbook}.
Recently, considerable research has been done in the direction of network spectra
\cite{spectra-net,Aguiar-SF}. 

In the following, we mention known results on the spectra of real world and model networks.
The spectra of networks have some correspondence with the spectra of random 
matrices. For instance, the distribution of eigenvalues of a matrix having finite mean 
number $p$ of nonzero Gaussian distributed random elements per row follows Wigner 
semicircular law in the limit $p \rightarrow N$, where $N$ is the dimension of matrix 
\cite{mehta,rev-rmt}. For very small 
$p$, which corresponds to the sparse random matrix, one gets the semicircular law 
but with peaks at different parts of the spectrum (maximum at the eigenvalue {\it zero}) \cite{sparseRM}. Recent 
investigations of the spectral behavior of networks, leading to matrices with entries $zero$ and 
$one$, show that the random networks \cite{erdos} follow Wigner semicircular law 
as well \cite{Vicsek} with degeneracy at the eigenvalue {\it zero}. The small-world model networks \cite{SW} 
show a very complex spectral density with many sharp peaks \cite{SJ_pre2007a}, while the 
spectral density of the scale-free model networks \cite{BA} exhibits a triangular
distribution \cite{Vicsek,Aguiar-SF,Chung,SJ_pre2007a}. The spectra of real world networks show remarkably different
features than that of the model networks \cite{Aguiar-SF,Chung,SJ_pre2007a,Jost}, 
and based on this observation  
a network construction method was proposed which captures a peak at zero property shown by the spectra of
many real world networks such as protein-protein interaction networks \cite{Jost}. 
Recently, spacing distributions of
Erd\"os-R\'enyi networks have been studied under random matrix theory (RMT) framework 
\cite{Vattay}. As connection probability decreases
Ref.~\cite{Vattay} shows a transition to the Poisson statistics. Additionally, it shows
the transition to the Poisson statistics upon the deletion of nodes in the 
real world networks \cite{Vattay}. 
Refs.~\cite{SJ_pre2007a,SJ_pre2007b}  
have shown that the spacing distributions of 
various model networks, namely small-world and scalefree networks,
follow the universal behavior of RMT. In contrast to \cite{Vattay}, these works \cite{SJ_pre2007a,SJ_pre2007b} have considered only connected networks.
Furthermore, spectral rigidity such as the $\Delta_3$ statistics, defined in 
Eq.~\ref{delta3}, provides a qualitative measure of the 
level of randomness in networks \cite{SJ_new}. Recently localization 
of eigenvectors have also been used to analyze various structural and dynamical properties of 
real and model networks \cite{eigenvector}. 

RMT, initially proposed to explain statistical properties of
nuclear spectra, has also provided successful predictions for the spectral properties of different
complex systems such as disordered systems, quantum chaotic systems and large complex
atoms among this. It has been followed by numerical and experimental verifications in the last few
decades \cite{mehta,rev-rmt}. Quantum graphs, which model the systems of interest in
quantum chemistry, solid state physics and transmission of waves, have also been
studied under the RMT framework \cite{Qgraph}. Recently, RMT has been shown to be
useful in understanding the statistical properties of empirical
cross-correlation matrices appearing in the study of multivariate time series in
several problems
: price fluctuations in stock market \cite{rmt-stock}, Electro encephalogram
data \cite{rmt-brain}, variation of different atmospheric parameters
\cite{rmt-atmosphere}.

In the present work we study spectral behavior of networks having 
community structure under the framework of RMT. The study of community structure helps to elucidate 
the organization of networks, and eventually could be related to the functionality of groups
of nodes \cite{Amaral, Newman,community}. 
Regardless of the type of real world networks in terms of the degree and 
other structural properties \cite{rev-network}, it is possible to distinguish communities in the
whole networks \cite{Newman}. However, the question of definition of the community is
problematic, and usually community is assigned to the nodes which are connected densely
among themselves, and are only sparsely connected with other nodes outside the
community. We therefore model here community structure by sparsely connected Erd\"os-R\'enyi 
random networks. This simple approach considers more densely connected nodes as a 
definition of community, and does not pay attention to the detailed structure of the
connections \cite{Amaral}. 
Recent literature is largely filled up with methods 
to detect communities in
networks based on structural measures \cite{Vespignani,comm-det}, whereas few works 
emphasize on the spectral properties such as density distribution and eigenvector analysis as well \cite{comm-det-spectra}. 
The objective of our work is not the detection of communities, rather we show the applicability
of spectral methods under the RMT framework to analyze community structures in networks. Instead of
paying attention to the nodes forming communities, we look for the signatures of overlapping of communities
in the spectra of the corresponding adjacency matrix. 
We study various spectral behaviors, namely density 
distribution, nearest neighbor spacing distribution (NNSD) and spectral rigidity for deformed 
random networks. 
We find that the NNSD
detects even the small mixing of communities in the network, whereas spectral rigidity probed
by the $\Delta_3$
statistics is suitable to analyze larger mixing, which is, in general, the case for real world
networks. Communities are modeled by random or scale-free sub-networks, and interactions between
communities are considered as random. For small interaction strength the NNSD of the network 
shows the transition from the Poisson to the Gaussian Orthogonal Ensemble (GOE) statistics. 
For large interactions, the $\Delta_3$
statistics shows systematic increase in the range for which it follows GOE statistics. 
Finally, as an application, we study the spectral properties of a protein-protein interaction
network of Helicobacter under the RMT framework.

\section{Deformed networks}
For an unweighted network, the adjacency matrix is defined in the following way
: $A_{ij} = 1$, if $i$ and $j$ nodes are connected and $zero$ otherwise. For undirected
networks, this matrix is symmetric and consequently has real eigenvalues. Random 
matrices corresponding to unweighted random networks have entries $0$ and $1$, where 
number of $1$'s in a row follows a Gaussian distribution with mean $p$ and variance 
$p(1-p)$. This type of matrix is very well studied within the RMT 
framework \cite{mehta,sparseRM}. We then turn our attention to the following structure: 
(1) Take $m$ random networks with connection probability $p$; the spectral behavior of 
the matrix corresponding to each of these sub-networks (blocks) separately follows 
GOE statistics. The matrix corresponding to the full network would be a $m$ block diagonal 
matrix. (2) Introduce random connections among these sub-networks with probability $q$. This configuration 
leads to $m$ block matrix, with blocks having entries $one$ with portability $p$, and
off diagonal blocks having entries $one$ with probability $q$. The above 
networks can be casted in the following form:
\begin{equation}
A = A_0 + A_q
\label{net}
\end{equation} 
\begin{figure}
\centerline{\includegraphics[width=\columnwidth]{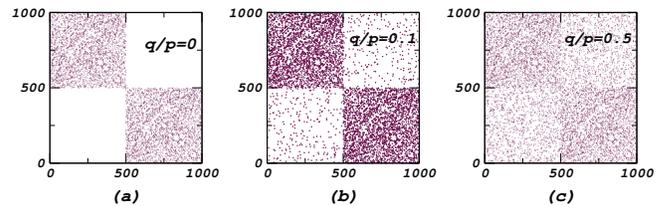}}
\caption{(color online) Connection matrices corresponding to $p=0.01$ and 
different values of $q$. 
(a) plots the connection matrix of the two sub-networks which do not have any 
connection between them. (b) corresponds to $q/p=0.1$, and (c) depicts the case 
$q/p=0.5$, when the connections between the the sub-networks are as large as $50\%$ of the 
connections inside.}
\label{Fig_Net}
\end{figure}
$A_0$ is a $m$ blocks diagonal random matrix, where each block represents one community, and the
off-diagonal block matrix $A_q$ denotes the interactions among the communities. Each 
block in $A_q$ is a random matrix, which for large $N$ has mean $q$ and deviation $q(1-q)$. 
Since the nonzero values of $q$ introduce deformation to the complete block diagonal form,
we refer $A$ being a deformed network. This terminology is motivated by the literature on
deformed random matrices \cite{deformed}.
Fig.~\ref{Fig_Net} shows the connection matrices for $m=2$ and various values of 
$q$. Fig.~\ref{Fig_Net}(a) represents the two random sub-networks, each of size $N=500$, with the
connection probability inside a sub-network being $p=0.01$ and between the sub-networks 
being $q=0$. The ratio $q/p$, which can be considered as the relative strength of $A_q$ and 
$A_0$, measures the deformation from the block-diagonal form of the matrix, or from the 
perfect structured network. The value $q/p=1$, which corresponds to equal strength of
inter and intra-community connections, yields complete random network. 
Fig.~\ref{Fig_Net}(b) plots 
the connection matrix for $q/p=0.1$, which implies that inter-community connections
are $10\%$ of the intra-community connections. Fig.~\ref{Fig_Net}(c) shows the connection matrix for $q=0.005$; 
for this value of $q$, the inter-community strength  
is $50\%$ ($q/p=0.5$) of the intra-
community strength. Note that in numerical 
simulations we use the value of $p$ equal to 0.01, which leads to a sparse
connected random network ($N_c \sim N$) with the average degree $ <k> \sim N \times p = 5$,
$N_c$ being the number of connections in the network. 
Larger value of $p$ would lead to networks with the larger average degree. Real 
world networks are sparse \cite{rev-network}, and hence we chose such a small value of $p$.   

\section{Numerical Simulation Results}
\begin{figure}
\centerline{\includegraphics[width=\columnwidth]{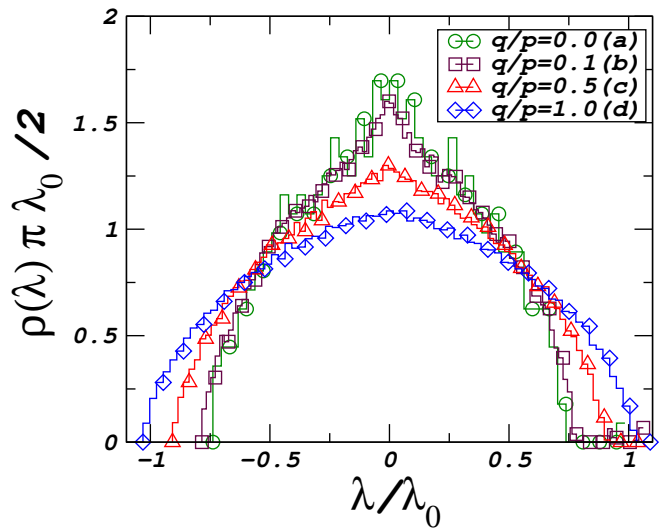}}
\caption{(color online) Density distribution of the two random sub-networks connecting 
each other with probability (a) $q=0$ and $q/p=0$; (b) $q=0.001$ and hence $q/p=0.1$;
(c) $q=0.005$, hence $q/p=0.5$ and (d) $q/p=1$ which corresponds to $q=0.01$ . 
Each block (random network) has size $N=500$. The axes are scaled in such a way that the
semicircle corresponding to $q=p$ has unit radius (see text). All
graphs are plotted for 20 realizations of random sets of connections among the two
sub-networks.}
\label{Fig_Density}
\end{figure}

We denote the eigenvalues of the network by $\lambda_i,\,\,i=1,\dots,m \times N$, where $N$ 
is the size of the sub-network, and $m$ is the number of the sub-networks. Note that the size
of each sub-network may be different, but for simplicity we consider here equal
size.
Fig.~\ref{Fig_Density} plots the spectral density for $m=2$ block matrices having 
$q N^2$ non-zero off diagonal entries, corresponding to the two sub-networks connected with 
probability $q$. As discussed earlier $q$ varies from $q=0$, which corresponds to the
two completely disconnected 
sub-networks ( $A=A_0$, Fig.~\ref{Fig_Net}(a) ), to $q=p$ leading to a single random 
network. The cases for $0 < q << p$ correspond to the configurations when the 
initial community structure is almost preserved. Increase in the value of $q$ leads 
more entries of $one$ in the 
matrix $A_q$ (Eq.~\ref{net}). Finally the $q=p$ case destroys the community structure 
completely, and the network 
can be treated as one single random network. Fig.~\ref{Fig_Density} presents the density 
distribution of eigenvalues for various values of $q$. The eigenvalues are scaled with respect to the 
spectra of the network for $q/p=1$. With this scaling, the density distributions are not 
semicircular for values of $q < p$. As the coupling between the two blocks increases ($q>0$), 
the density distribution shows a transition to the semicircular form at $q=p$:
\begin{equation}
\rho(\lambda) = \frac{2}{\pi \lambda_0^2} \sqrt{(\lambda_0^2 - \lambda^2)},
\nonumber
\end{equation}
where $\lambda_0$ is the radius of the semicircular distribution for $q=p$ calculated from the
spectra of network as $\lambda_0 = (\lambda_{max} - \lambda_{min})/2$, 
$\lambda_{max}$ and $\lambda_{min}$ being the highest and the lowest eigenvalue. Now we turn
our attention to the statistics of eigenvalue fluctuations.
\subsection{Nearest neighbor spacing distribution}
\begin{figure}
\centerline{\includegraphics[width=\columnwidth]{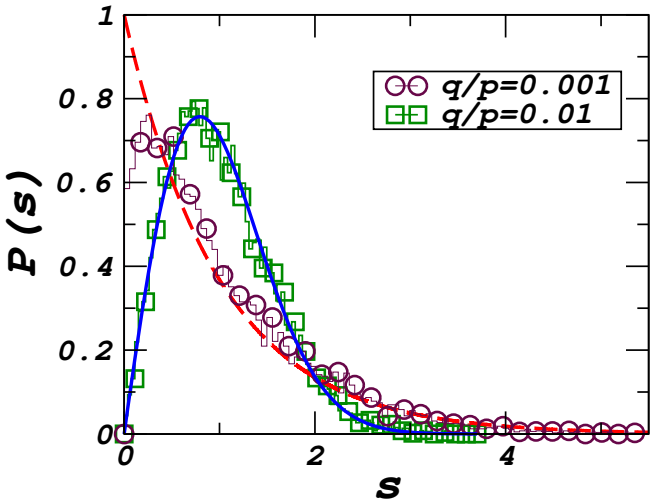}}
\caption{(color online) Nearest neighbor spacing distribution for the two such 
values of $q$ which gives
two extreme statistics. Histograms correspond to
the numerical values $q=10^{-5} (q/p=0.001)$ and $q \ge 10^{-4} (q/p \ge 0.01)$.
The solid and dotted curves are, respectively, Poisson and GOE predictions of
RMT. The figure is plotted for an average over 20 realizations of the random set of connections
between the networks.}
\label{Fig_Spac}
\end{figure}

In the following, we study spectral fluctuations of the networks for different
values of $q$. In order to get universal properties of the eigenvalue
fluctuations, one has to remove the spurious effects due to variations of the
spectral density and to work at the constant spectral density on the average. Thereby,
it is customary in RMT to unfold the eigenvalues by a transformation
$\overline{\lambda}_i = \overline{N} (\lambda_i)$, where $\overline{N} (\lambda) =
\int_{\lambda_{\mbox{\tiny min}}}^\lambda\, \rho(\lambda^\prime)\, d
\lambda^\prime$ is the averaged integrated eigenvalue density \cite{mehta}.
Unfolding is a transformation which produces  the eigenvalues with a constant
average level density. Since we do not have an analytical form for $\overline{N}$, we
numerically unfold the spectrum by polynomial curve fitting.

Using the unfolded eigenvalues, we calculate the NNSD $P(s)$, 
where $s^{(i)}=\overline{\lambda}_{i+1}-\overline{\lambda}_i$, for different $q$ values. 
Fig.~\ref{Fig_Spac} plots the spacing distribution for the
two values of $q$, $q=0$ and $q=10^{-4}$. For such small values of $q$, 
although the density distributions remain unchanged, the NNSD shows
significant changes. Spacing distributions calculated from the network spectra are
fitted using Brody formula \cite{Brody},
\begin{equation}
P_{\beta}(s) = A s^\beta\exp\left(-\alpha s^{\beta+1}\right),
\label{brody}
\end{equation}
where $A$ and $\alpha$ are determined by the parameter $\beta$ as follows :
\begin{equation}
A\,=\,(1+\beta) \alpha\,\mbox{and}\,\alpha = \left[\Gamma\left(\frac{\beta+2}
{\beta+1}\right)\right]^{\beta+1}.
\nonumber
\end{equation}
Eq.~\ref{brody} is a semi-empirical formula characterized by the single
parameter $\beta$. As $\beta$ goes from zero to one, the Brody formula smoothly
changes from Poisson to GOE. As can be seen from Fig.~\ref{Fig_Spac}, for $q/p \sim 0.001
(q \sim 10^{-5})$, 
the value of the Brody parameter $\beta
\sim 0.2$, which suggests that distribution is very close to the Poisson
[$P(s)=\exp(-s)$] denoted by the dotted curve in the figure. As the value of
$q$ increases, $\beta$ also increases, and it is of the order of 1 for the
value of $q/p \sim 0.01$ (which corresponds to the value of $q$ as less as
$10^{-4}$), and becomes insensitive for a further increase in $q$. For
larger values of $q$, we analyze the spectra using the spectral rigidity test of RMT.

\subsection{Spectral rigidity via $\Delta_3$ statistics}
\begin{figure}
\centerline{\includegraphics[width=\columnwidth]{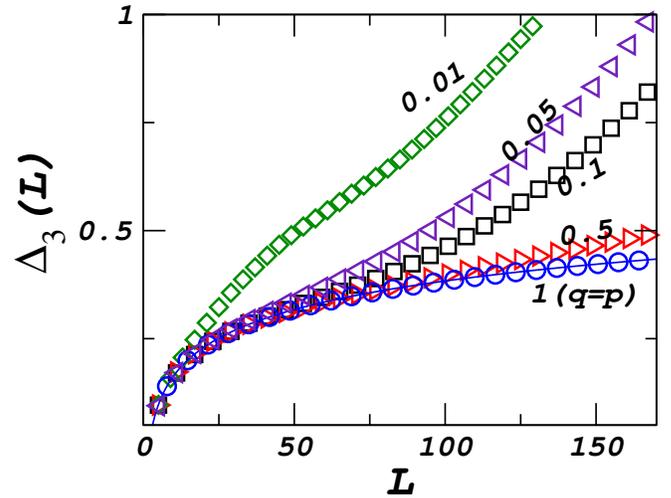}}
\caption{(color online) Long-range correlations among eigenvalues. Different open 
symbols 
are the numerical values of $\Delta_3$ for various $q$ values, and the 
solid curve (merged with the open circles corresponding to $q/p=1$)
is the GOE prediction (Eq.~\ref{delta3}). Since for $q=p$ the $\Delta_3$ statistics of 
network follows the GOE prediction completely, the solid line showing GOE statistics merges with the circles showing
numerical values for this $q$. The figure is plotted for an average over 20 
realizations of 
the networks. $\Delta_3$ follows the universal RMT prediction up to certain $L$ values. The 
range of $L$ for which $\Delta_3$ follows GOE statistics increases with the ratio 
$q/p$. }
\label{Fig_Delta3}
\end{figure}
The spectral rigidity, measured by $\Delta_3$ statistics of RMT, gives 
information about the long-range correlations among the eigenvalues. The 
$\Delta_3$ statistics measures the least-square deviation of the spectral staircase 
function representing the cumulative density $N(\overline{\lambda})$ from the best 
straight line fitting for a finite interval $L$ of the spectrum, i.e.,
\begin{equation}
\Delta_3(L; x) = \frac{1}{L} \min_{c_1,c_2} \int_x^{x+L} \,\left[
N(\overline{\lambda}) - c_1 \overline{\lambda} -c_2 \right]^2\,d \overline{\lambda}
\label{delta3}
\end{equation}
where $c_1$ and $c_2$ are obtained from a least-square fit. Average over several 
choices of $x$ gives the spectral rigidity $\Delta_3(L)$. 
For the uncorrelated eigenvalues, $\Delta_3 (L) = L/15$, reflecting 
strong fluctuations around the spectral density $\rho(\lambda)$. For the
GOE case, $\Delta_3(L)$ statistics is given by  
\begin{equation}
\Delta_3(L) \sim \frac{1}{\pi^2} \ln L .
\label{delta3_goe}
\end{equation}
Fig.~\ref{Fig_Delta3} plots the $\Delta_3$ statistics for five different values of 
$q$. Various open symbols are the numerical values of $\Delta_3$ for various $q$ values,
and the solid line (merged with the $q/p=1$ case) is the $\Delta_3(L)$ statistics for the GOE case (Eq.~\ref{delta3_goe}).
As seen from Fig.~\ref{Fig_Delta3}, the $\Delta_3(L)$ 
statistics follows RMT predictions of GOE (Eq.~\ref{delta3_goe}) up to a certain $L$. It has
a linear behavior in semi-logarithmic scale 
with the slope of $\sim 1/\pi^2$. The value of $L$ for which 
it follows GOE statistics depends upon $q$. For small values of $q$ such as 
$q/p=0.01$ and $q/p=0.05$, $\Delta_3$ follows RMT prediction till very 
small range of $L \sim 5$ and $L \sim 20$, respectively. As $q$ increases, the 
value of $L$ for which $\Delta_3$ follows the GOE statistics also increases. 
For $q/p=0.1$, it agrees with the RMT predictions of 
GOE behavior for 
$L \sim 75$, and after this value, deviation from the RMT prediction is seen. This  
deviation corresponds to the existence of community structure in the network. As the value of 
$q$ increases, the communities have more and more random connections between them. For 
$q=p$ the community structure is destroyed fully, and the network is a complete 
random network. This fact is reflected in the $\Delta_3$ statistics corresponding 
to $q/p=1$. At this value of $q$, it follows RMT prediction up to a 
very long-range $L \sim 150$. After this value of $L$, for the network of size 
$N \times m=1000$ we do not have a meaningful calculation of the $\Delta_3$ statistics 
\cite{casati}. For $q=0.005$ $(q/p=0.5)$ (see Fig.~\ref{Fig_Net}), where 
the strength of inter-community is as large as 
$\sim 50\%$ of the intra-community connections strength, the $\Delta_3$ statistics correctly reflects 
the deviation from complete random matrices, suggesting the existence of communities in the 
network.

Note that we present results for each sub-network having equal size.  For sub-networks
having different sizes, all the figures remain the same. The crucial quantity, which affects 
correlations
of eigenvalues, is the variance of each block or the ratio $q/p$. For blocks
with different sizes, but with the same $q/p$, similar results are obtained, 
except for the exact value of $L$ in Fig.~\ref{Fig_Delta3} for which the $\Delta_3$ 
statistics follows GOE distribution, which scales with the network size \cite{SJ_new}.

\begin{figure}
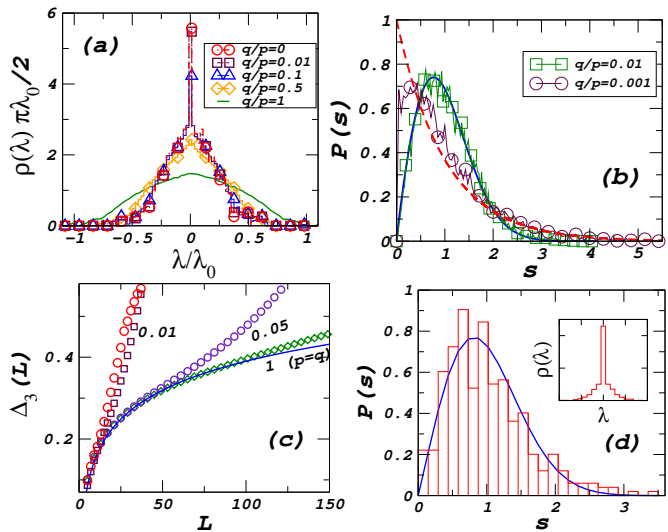

\centerline{\includegraphics[width=0.48\columnwidth]{fig5a.eps}
\includegraphics[width=0.52\columnwidth]{fig5b.eps}}
\centerline{\includegraphics[width=0.52\columnwidth]{fig5c.eps}
\includegraphics[width=0.48\columnwidth]{fig5d.eps}}
\caption{(color online) Spectral behavior of deformed scalefree networks. Sub-networks
are scalefree networks of size $N=500$ and average degree $<k> = 5$. (a) plots
the eigenvalue density distribution of deformed scalefree network for various values of $q/p$. 
(b) and (c) plot the
NNSD and $\Delta_3$ statistics respectively. Dashed and solid line in (b) corresponds to 
Poisson and GOE statistics respectively. All graphs are plotted 
for 20 sets of random realization of interaction networks. (d) plots the density 
distribution (inset), and spacing distribution for a protein-protein interaction network in 
Helicobacter as histogram and GOE statistics as solid line. This network has size $N=712$ and average degree $<k> \sim 5$. 
Open circles in (c) is the $\Delta_3$ statistics for Helicobacter.}
\label{Fig_Scale}
\end{figure}
\section{Deformed scalefree networks}
In the following we consider scalefree networks as the sub-networks, and study 
the spectral behavior for various values of $q$. Again $q$ measures the strength of 
the off-diagonal block matrix defining the interaction between the sub-networks. 
Matrix $A_0$ in  Eq.~\ref{net}, corresponding to the scalefree sub-networks, consists of two block 
diagonal matrices, with entries of one in each block following a power law characteristic 
of the sub-network.  We use Barab\'asi-Albert algorithm \cite{BA} to generate 
the scalefree sub-networks. In scalefree network the probability $P(k)$, that a node has 
degree $k$, decays as a power law $P(k) 
\sim k^{-\gamma}$, where $\gamma$ is a constant and for the type of probability
law used in the simulations $\gamma = 3$. Other forms for the probability law are also 
possible which gives different exponent \cite{BA-exact}. However, the results reported here are independent of the 
value of $\gamma$ \cite{note1}. Size and average degree of the sub-networks remain the 
same as for the random sub-networks, 
i.e. $N=500$ and $<k>=5$.  The average degree ($<k>$) of a network can be
calculated as $<k>=2 \times N_c/N$, where $N_c$ is the number of connections and $N$ is the size
of the network. With the increase in the value of $q$, deformation from the 
network having scalefree community structure also increases. Fig.~\ref{Fig_Scale} plots 
various spectral behavior of deformed networks made of the scalefree sub-networks. 
Fig.\ref{Fig_Scale}(a) plots the density 
distribution for the various values of $q$. For small values of $q$, the density 
is very different from that of the deformed random networks (Fig.~\ref{Fig_Density}). 
It has a triangular shape with a peak at zero. This is a well-known shape for sparse
scalefree 
networks \cite{Aguiar-SF,SJ_pre2007a,Chung}. For $q/p < 0.01$, when the scalefree 
structure of the sub-networks dominates over the random interaction between them, the 
eigenvalue density distribution does not show any noticeable change. But the NNSD
in Fig.~\ref{Fig_Scale}(b) suggests a possible structure in the network. 
As shown in 
Fig.~\ref{Fig_Scale}(b), for $q/p=0.001$ ($q = 10^{-5}$) the NNSD is close to 
Poisson statistics 
with a value of the Brody parameter $\beta \sim 0.21$. As $q$ increases, value 
of the Brody parameter increases as well, becoming one for $q \sim 10^{-4}$. After 
this value of $q$, the NNSD does not provide any further insight, and we probe
for long-range correlations among eigenvalues. Fig.~\ref{Fig_Scale}(c) plots 
the $\Delta_3$ statistics for various values of $q$. It shows similar behavior as for 
the deformed random networks (see Fig.~\ref{Fig_Delta3}). For $q \sim 0.01$, 
when the network has distinguishable community structure, the value of $L$ for which 
$\Delta_3$ follows the GOE statistics (\ref{delta3_goe}) is as small as 25. 
As $q$ is increased, 
$L$ also increases, becoming $\sim 150$ for $q/p \sim 1$. 

Fig.~\ref{Fig_Scale}(d) shows the density distribution (inset) and the 
spacing distribution of 
the protein-protein interaction network of Helicobacter \cite{cosin}. The largest 
connected component of the network has dimension $N=708$ and number of connections 
$N_c = 2789$. The average degree of this scalefree network is $<k> \sim 4$. The density 
distribution has triangular form with a peak at zero. This behavior of the density distribution 
suggests scalefree properties of the network \cite{Aguiar-SF,SJ_pre2007a,Chung}, but does not 
provide information of randomness or structure in the network. To get further insight, we 
calculate the NNSD and the spectral rigidity of the network. For this, first we unfold the eigenvalues 
using the procedure explained earlier. The NNSD of the network follows GOE statistics with 
the value of $\beta \sim 0.98$, suggesting enough random connections in the network.
Further test of long-range correlations among eigenvalues shows that the $\Delta_3$ statistics 
follows the GOE prediction (Eq.~\ref{delta3_goe}) up to $L \sim 20$ in 
Fig.~\ref{Fig_Scale}(c), 
and after this value 
deviation from the universal behavior is seen. It suggests that, though the network has enough random 
connections which give rise to short-range correlations among eigenvalues, it
has strong community structure causing deviation of the $\Delta_3$ statistics 
from the random matrix behavior after a certain range.

In the present paper we consider only the random interactions between communities. For other kind of
interactions, for instance interactions among the scalefree sub-networks as considered 
in \cite{Amaral} which leads to a hierarchical scalefree network, the density distribution
would show an entirely different behavior from the semi-circular distribution.
$p=q$ case would lead to a scalefree topology which has a triangular density distribution with 
peak at zero. However,
spectral fluctuations would show qualitative similar behavior. For small coupling interactions
among the sub-networks, the NNSD results would be same as presented here, showing a
transition from Poisson to GOE statistics \cite{SJ_new}, whereas for
large coupling interactions the exact range for which the $\Delta_3$ statistics follows GOE 
would be different from those of the random interactions. Further detailed results 
of this model as well as real world networks having more complicated structures analyzed 
under the deformed random matrix framework would be discussed elsewhere 
\cite{SJ_Mahir_new}.
 
\section{Conclusions and Discussions}
The eigenvalue density distribution of networks having two sub-networks tend towards 
the semi-circular distribution as the random connections between the sub-networks are increased. 
For very small values of $q < 10^{-4}$, corresponding to the very small 
deformation from the community structure, the density distribution does not present any 
noticeable changes, but the NNSD, which reflects short-range correlations among eigenvalues, 
show important features.  For two 
random sub-networks, which are almost uncoupled (i.e. $q \sim 0$), the NNSD is very close to the 
Poisson statistics, and as $q$ increases, it has a smooth transition to the GOE statistics. 
Note that this Poisson to GOE transition is found for many different systems, for example
spectra of insulator-metal transition, order-chaos transition  
follow this Poisson-GOE transition  \cite{rev-rmt}. Sade et. al \cite{Sade} 
have studied transition to the GOE statistics as a function of site disorder for the spectra 
of small-world and scale-free networks. Here, by keeping the network structure fixed, disorder at 
nodes is increased and depending upon the network average degree transition to GOE statistics is
seen. The main difference between \cite{Sade} and the study presented in this paper is the following: we track
changes in the spectra with structural changes in the network architecture. As random
connections among the sub-networks are increased, first there is transition for the NNSD to the GOE statistics, and 
this transition occurs for very small value of random connections among networks. This
is the crucial and remarkably different result observed here, which suggests that
very small random interaction between communities is enough to introduce short-range
correlations among them, spreading the randomness in the whole network. Second,
further increase in coupling among the sub-networks is reflected by long-range correlations
among eigenvalues.
For this increase in the value of $q$, the NNSD does not give additional insight to the deformation of the network, 
as it 
remains same with the $\beta \sim 1$, so we turned our attention to the $\Delta_3$ statistics.

The $\Delta_3$ statistics, which measures long-range correlations among the eigenvalues,
detects deformation from a network having two coupled sub-networks,
to a single random network. More deformation of the network from community structure, leads to
a larger range of $L$ for which $\Delta_3$ follows the GOE statistics. Note that, for
the case of sub-networks being completely random, the spacing and the $\Delta_3$ statistics
of each of them follows RMT prediction. Therefore,
any deviation from GOE statistics is due to the community structure these two sub-networks form
when considered as a single network.

It is interesting to note that our results resemble the behavior of deformed
random matrix ensembles (DGOE) introduced to study the effect of isospin symmetry breaking
in nuclei \cite{deformed}. The qualitative behavior of the spectral density and 
the $\Delta_3$ statistics of networks presented here is similar to that of deformed 
matrices studied in \cite{Mahir,Mahir_density,Mahir_delta3}. The analytical
form of the density derived in \cite{Mahir_density} depends on a parameter $\alpha$
measuring the relative strength of the off-diagonal random matrices to the block
diagonal random matrices. In similar lines, for deformed networks,
we can compare $q/p$, relative strength of off-diagonal and diagonal networks,
with $\alpha$.  The results presented 
here suggest that further investigations of complex networks following similar 
lines as in deformed random matrices \cite{Mahir_density} would be useful to have detailed 
information of communities in the networks \cite{SJ_Mahir}.

To conclude, we have studied the spectral behavior of networks having community structure,
and shown that the NNSD and $\Delta_3$ statistics capture features related to the
structure in the network. We investigate the spectral properties of
a real world network as well, and compare the results with those of the model networks.
On the one hand, results presented in this paper advances the studies of the spectral properties
of network with the community structure under the universal RMT
framework; on the other hand, variations in the
correlations among eigenvalues shed light on the coupling among communities.
For the simulations, the community structure in
network is modeled by the very simple random or scalefree sub-networks, and the interactions among
these sub-networks are considered random, whereas real world networks have 
richer structure \cite{Amaral}. However, the results presented here provide a platform
to investigate the community structure of networks using a well developed theory of random
matrices; the further investigations in this direction would deal with real world
networks with richer and more complicated structure under the 
deformed random matrix framework \cite{SJ_Mahir_new,SJ_Mahir}. 

\section{Acknowledgments} We acknowledge Dr. M. Hussein for
useful suggestions and Dr. G. Vattay for stimulating discussions about prospective of the results.


\begin{thebibliography}{99}
\bibitem{rev-network} R. Albert and A.-L. Barab\'asi, Rev. Mod. Phys.
{\bf 74}, 47 (2002) and references therein; S. Boccaletti, V. Latora, Y. Moreno, M. Chavez,
D.-U. Hwang, Phys. Rep. {\bf 424}, 175 (2006).

\bibitem{SW} D. J. Watts and S. H. Strogatz, Nature {\bf 440}, 393 (1998).

\bibitem{BA} A.-L. Barab\'asi and R. Albert, Science {\bf 286}, 509 (1999).

\bibitem{Newman} M. Girvan and M. E. J. Newman, Proc. Natl. Acad. Sci. USA {\bf 99},
7821 (2002); M. E. J. Newman, Social Networks {\bf 27}, 39 (2005);
M. E. J. Newman, Proc. Natl. Acad. Sci. USA {\bf 103}, 8577 (2006); M. J. Krawczyk,
Phys. Rev. E {\bf 77} 065701 (R) (2008).

\bibitem{Amaral} E. Ravsaz {\it et al.}, Science {\bf 297}, 1551 (2002);
R. Guimer\'a and L. A. N. Amaral, Nature {\bf 433}, 895 (2005).

\bibitem{spectrum} D. M. Cvetkovi\'c, M. Doob and H. Sachs,
{\it Spectra of Graphs : theory and applications}, (Academic Press,
3rd Revised edition, 1997).

\bibitem{handbook} M. Doob in {\it Handbook of Graph Theory}, edited by
J. L. Gross and J. Yellen (Chapman \& Hall/CRC, 2004).

\bibitem{spectra-net} K. -I. Goh, B. Kahng, and D. Kim, Phys. Rev. E {\bf 64}, 051903 (2001);
S. N. Dorogovtsev, A. V. Goltsev, J. F. F. Mendes and A. N.
Samukhin, {\it ibid} {\bf 68}, 046109 (2003); E. Estrada, Europhys. Lett. {\bf 73} (4)
649 (2006); D. Kim and B. Kahng, Chaos {\bf 17} 026115 (2007); H. Yang, C. Yin, 
G. Zhu and B. Li, Phys. Rev. E {\bf 77} 045101(R) (2008);  G. Bianconi, arXiv:0804.1744.

\bibitem{Aguiar-SF} M. A. M. de Aguiar and Y. Bar-Yam, Phys. Rev. E
{\bf 71}, 016106 (2005); A. N. Samukhin, S. N. Dorogovtsev, and J. F. F. Mendes
{\it ibid} {\bf 77}, 036115 (2008).

\bibitem{mehta} M. L. Mehta, {\it Random Matrices}, 3rd ed. (Elsevier Academic
Press, Amsterdam, 2004).

\bibitem{rev-rmt} T. Guhr, A. Muller-Groeling and H. A. Weidenmuller,
Phys. Rep. {\bf 299}, 189 (1998).

\bibitem{sparseRM} S. N. Evangelou, Journal of Statistical Physics, {\bf 69}, 361
(1992).

\bibitem{erdos} P. Erd\"os and A. R\'enyi, Publ. Math. Inst. Hungar.
Acad. Sci. {\bf 5}, 17 (1960).

\bibitem{Vicsek} I. J. Farkas, I. Der\'enyi, A. -L. Barab\'asi, and T. Vicsek,
Phys. Rev. E {\bf 64}, 026704, (2001).

\bibitem{SJ_pre2007a} J. N. Bandyopadhyay and S. Jalan, Phys. Rev. E {\bf 76}, 026109  (2007).

\bibitem{Chung} F. Chung, L. Lu and V. Vu, Proc. Natl. Acad. Sci. USA {\bf 100}
6313 (2003).
\bibitem{Jost} A. Banerjee and J. Jost,
Networks and Heterogeneous Media {\bf 3}, 395 (2008); A. Banerjee and J. Jost,
Lin. Alg. Appl., {\bf 428}, 2008, 3015-3022;

\bibitem{Vattay}G. Palla and G. Vattay, New J. Phys. {\bf 8} 307 (2006).

\bibitem{SJ_pre2007b} S. Jalan and J. N. Bandyopadhyay, Phys. Rev. E {\bf 76}, 046107 (2007).
%\bibitem{SJ_2008} S. Jalan and J. N. Bandyopadhyay, Physica A {\bf 387}, 667 (2008).

\bibitem{SJ_new} S. Jalan and J. N. Bandyopadhyay, Euro. Phys. Letts. ({\it in press}). 

\bibitem{eigenvector} P. N. McGraw and M. Menzinger,
Phys. Rev. E {\bf 77} 031102 (2008); G. Zhu, H. Yang, C. Yin and B. Li, {\it ibid} 
{\bf 77}, 066113 (2008).

\bibitem{Qgraph}T. Kottos and U. Smilansky, Phys. Rev. Lett. {\bf 79}, 4794 (1997);
J. Phys. A: Math. Gen. {\bf 36}, 3501 (2003).

\bibitem{rmt-stock} L. Laloux, P. Cizeau, J.-P. Bouchaud, and M. Potters,
Phys. Rev. Lett. {\bf 83}, 1467 (1999); V. Plerou, P. Gopikrishnan, B. Rosenow,
L. A. N. Amaral, and H. E. Stanley, Phys. Rev. Lett.
{\bf 83}, 1471 (1999).
\bibitem{rmt-brain} P. Seba, Phys. Rev. Lett. {\bf 91}, 198104 (2003).

\bibitem{rmt-atmosphere} M. S. Santhanam and P. K. Patra, Phys. Rev. E {\bf 64}, 016102 (2001).

\bibitem{community}G. Palla, I. Derenyi, I. Farkas and T. Vicsek, Nature 435, 814 (2005);
M. E. J. Newman, Phys. Rev. E {\bf 70}, 056131 (2004); A. Arenas, A. Fernandez and S. Gomez, New J. Phys.
{\bf 10} 053039 (2008).

\bibitem{Vespignani}V. Colizza, A. Flammini, M. A. Serrano and
A. Vespignani,  Nature Physics {\bf 2}, 110 (2006).

\bibitem{comm-det}M. E. J. Newman, Phys. Rev. E {\bf 69}, 066133 (2004); M. B. Hastings
Phys. Rev. E {\bf 74}, 035102 (2006); S. Fortunato and M. Barthelemy,
Proc. Natl. Acad. Sci. USA {\bf 104}, 36 (2007) ; P. Schuetz and A. Caflisch,
Phys. Rev. E {\bf 78}, 026112 (2008). 

\bibitem{comm-det-spectra}M. E. J. Newman, Phys. Rev. E {\bf 74}, 036104 (2006) 

\bibitem{deformed}N. Rosenzweig and C. E. Porter, Phys. Rev. E {\bf 120}, 1698 (1960);
C. E. Porter, {\it Statistical Theory of Spectra : Fluctuations} (Academic, New York, 1965).

\bibitem{Brody} T. A. Brody, Lett. Nuovo Cimento {\bf 7}, 482 (1973).

\bibitem{casati} O. Bohigas, M. -J. Giannoni and C. Schmidt, in {\it
Chaotic behaviour in quantum systems} edited by G. Casati, p.103 (Plenum
Press, NewYork 1985).

\bibitem{BA-exact}S. N. Dorogovtsev, J. F. Mendes, and A. N. Samukhin, Phys. Rev. Lett. {\bf  85}, 
4633 (2000).

\bibitem{note1}The relation between the preferential attachment rule 
$\Pi(k) \propto k + a/m$
used in the manuscript and the degree distribution exponent $\gamma$ is,
$\gamma = 2 + a/m$ \cite{BA-exact}, where $a$ is the
initial attractiveness of nodes and $m$ is the number of connections new node makes.
For the Barab\'asi-Albert (BA) model \cite{BA}, $a/m=1$ which leads to the degree distribution
with $\gamma =3$. In the limit of zero initial connectivity $a=0$, all new nodes connect
only the first one. This case gives $\gamma=2$, and the network would be star network, with
$N-1$ nodes having $one$ connection and $one$ node with $N-1$ connections. The eigenvalues
of this star network are $-\sqrt{N-1}, 0, \sqrt{N-1}$, which gives the spectral density with
three peaks at these three values. The analysis and the comments about the spectral
behavior of scale-free networks presented in the section IV are made for
$2 < \gamma \le 3$. Most of the real world networks lie between
$2 < \gamma < 3$ \cite{rev-network}. For this range
the density distribution show typical triangular shape, and the tail of $\rho(\lambda)$
at large $\lambda$ is related to the behavior of the degree distribution $P(k)$.
In particular, as $P(k) \sim k^{-\gamma}$, $\rho(\lambda) \sim |\lambda|^{1-2\gamma}$ 
\cite{spectra-net}. Nearest neighbor spacing distribution for the individual sub-network 
would show GOE statistics of RMT \cite{SJ_pre2007b}.  The spectral behavior of the combined network would show
qualitative similar behavior of transition from Poisson to GOE statistics as coupling between
the sub-networks is increased, only the
range of $\Delta_3(L)$ statistics for which it follows GOE statistics may be different.  

\bibitem{cosin} http://pil.phys.uniroma1.it/~gcalda/cosinsite/extra/data/
proteins/helico

\bibitem{SJ_Mahir_new} J. X. de Carvalho, S. Jalan and M. S. Hussein ({\it under preparation}).
\bibitem{Sade}M. Sade and R. Berkovits, Phys. Rev. B {\bf 68}, 193102 (2003);M. Sade, T. Kalisky, S. Halvin
and R. Berkovits, Phys. Rev. E {\bf 72}, 066123 (2005).

\bibitem{Mahir}M. S. Hussein and M. P. Pato, Phys. Rev. Lett. {\bf 70}, 1089 (1993).

\bibitem{Mahir_density} A. C. Bertuola, J. X. de Carvalho, M. S. Hussein, M. P. Pato, and 
A. J. Sargeant, Phys. Rev. E {\bf 71} 036117 (2005)

\bibitem{Mahir_delta3}J. X. de Carvalho, M. S. Hussein, M. P. Pato and A. J. Sargeant, Phys.
Rev. E {\bf 76}, 066212 (2007).

\bibitem{SJ_Mahir} J. X. de Carvalho, S. Jalan and M. S. Hussein, Phys. Rev. E 
{\bf 79}, 056222 (2009)


\end{thebibliography}
\end{document}